\documentclass[10pt, conference]{IEEEtran}
\IEEEoverridecommandlockouts
\usepackage{cite}
\usepackage{amsmath, amssymb, amsfonts}
\usepackage{algorithmic}
\usepackage{graphicx}
\usepackage{textcomp}
\usepackage{xcolor}
\usepackage{hyperref} % Required for clickable links
\usepackage{booktabs}
\usepackage{tabularx}
\usepackage{array}
\usepackage{subcaption}
\usepackage{algorithm}
\usepackage{algorithmic}
\usepackage{multirow}

\usepackage{tcolorbox}

\newcommand{\findingbox}[2]{%
\begin{tcolorbox}[
  enhanced,
  width=\linewidth,
  colframe=black!65,
  arc=1pt,
  left=2pt,
  right=2pt,
  top=1pt,
  bottom=1pt,
  boxsep=1pt,
  before skip=2pt,
  after skip=2pt
]
\textbf{#1.} #2
\end{tcolorbox}%
}
\usepackage{listings}
\usepackage{lstlinebgrd}
\usepackage{float}

\definecolor{codebg}{RGB}{248,248,248}
\definecolor{codeframe}{RGB}{220,220,220}
\definecolor{codecomment}{RGB}{106,153,85}
\definecolor{codekeyword}{RGB}{0,92,197}
\definecolor{codestring}{RGB}{163,21,21}
\definecolor{callbg}{HTML}{E2EFFF}

\lstdefinelanguage{LLVMIR}{
  morekeywords={
    define,declare,global,constant,private,internal,external, dso_local,unnamed_addr,align,attributes,ret,br,switch,invoke, resume,unreachable,call,phi,select,icmp,fcmp,load,store, alloca,getelementptr,extractelement,insertelement,shufflevector,extractvalue,insertvalue,bitcast,addrspacecast,trunc,zext,sext,fptrunc,fpext,fptoui,fptosi,uitofp,sitofp,ptrtoint,inttoptr,add,fadd,sub,fsub,mul,fmul,udiv,sdiv,fdiv,urem,srem,frem, shl,lshr,ashr,and,or,xor,fneg, eq,ne,ugt,uge,ult,ule,sgt,sge,slt,sle, true,false,zeroinitializer,poison,null,undef, nsw,nuw,exact,inbounds,noundef,immarg, nocallback,nofree,nosync,nounwind,speculatable,willreturn, memory,none
  },
  sensitive=true,
  morecomment=[l]{;},
  alsoletter={\%@.\-_<>},
  morestring=[b]",
}

\lstdefinestyle{irstyle}{
  language=LLVMIR,
  basicstyle=\ttfamily\footnotesize,
  keywordstyle=\color{codekeyword}\bfseries,
  commentstyle=\color{codecomment}\itshape,
  stringstyle=\color{codestring},
  rulecolor=\color{codeframe},
  xleftmargin=6pt,
  xrightmargin=6pt,
  breaklines=true,
  breakatwhitespace=false,
  showstringspaces=false,
  columns=fullflexible,
  keepspaces=true,
  captionpos=b,
  numbers=none,
  tabsize=2
}

\lstdefinestyle{mystyle}{
  language=C++,
  basicstyle=\ttfamily\footnotesize,
  keywordstyle=\color{codekeyword}\bfseries,
  commentstyle=\color{codecomment}\itshape,
  stringstyle=\color{codestring},
  rulecolor=\color{codeframe},
  xleftmargin=6pt,
  xrightmargin=6pt,
  breaklines=true,
  breakatwhitespace=true,
  showstringspaces=false,
  columns=fullflexible,
  keepspaces=true,
  captionpos=b
}
\tcbuselibrary{skins,breakable,listings}

\definecolor{ragblue}{RGB}{54, 112, 180}
\definecolor{reasonorange}{RGB}{210, 126, 45}
\definecolor{patchgreen}{RGB}{64, 145, 108}
\definecolor{boxbg}{RGB}{248, 249, 250}
\definecolor{codebg}{RGB}{246, 248, 250}
\definecolor{bordergray}{RGB}{190, 190, 190}

\lstdefinelanguage{LLVMIR}{
  morekeywords={define,ret,icmp,and,or,xor,add,sub,mul,call,void,i1,i8,i16,i32,i64},
  sensitive=true,
  morecomment=[l]{;},
}

\lstdefinestyle{casecode}{
  basicstyle=\ttfamily\scriptsize\linespread{0.85}\selectfont,
  breaklines=true,
  columns=fullflexible,
  keepspaces=true,
  showstringspaces=false,
  frame=none,
  backgroundcolor=\color{codebg},
  xleftmargin=0.3em,
  xrightmargin=0.3em,
  aboveskip=0.1em,
  belowskip=0.1em,
  lineskip=-1pt
}

\newtcolorbox{casebox}[2][]{
  colback=boxbg,
  colframe=#2,
  coltitle=white,
  colbacktitle=#2,
  title=#1,
  fonttitle=\bfseries\scriptsize,
  boxrule=0.6pt,
  arc=1mm,
  left=0.6mm,
  right=0.6mm,
  top=0.6mm,
  bottom=0.6mm,
  boxsep=0.5mm
}
\newcount\DraftStatus  % 0 suppresses notes to selves in text
\newcommand{\nbc}[3]{\ifnum\DraftStatus=1
	{\colorbox{#3}{\bfseries\sffamily\scriptsize\textcolor{white}{#1}}}
	{\textcolor{#3}{\sf\small$\blacktriangleright$\emph{#2}$\blacktriangleleft$}}
\fi}

\newcommand{\draftnote}[2]{\ifnum\DraftStatus=1
	\marginpar{
		\tiny\raggedright
		\hbadness=10000
		\def\baselinestretch{0.8}
		\textcolor{#1}{\textsf{\hspace{0pt}#2}}}
\fi}

\def\BibTeX{{\rm B\kern-.05em{\sc i\kern-.025em b}\kern-.08em T\kern-.1667em\lower.7ex\hbox{E}\kern-.125emX}}
\begin{document}
\title{
Understanding Agent-Based Patching of \\Compiler Missed Optimizations
% Understanding and Improving Generalization-Aware Compiler Patching Agents
\\
% {\footnotesize \textsuperscript{*}Note: Sub-titles are not captured in Xplore and should not be used}
% \thanks{Identify applicable funding agency here. If none, delete this.}
}

\author{ \IEEEauthorblockN{Batu Guan}
\IEEEauthorblockA{\textit{The Chinese University of Hong Kong}\\ Hong Kong SAR \\ btguan@cse.cuhk.edu.hk}
\and
\IEEEauthorblockN{Zirui Wang}
\IEEEauthorblockA{\textit{The Chinese University of Hong Kong}\\ Hong Kong SAR \\ zrwang@cse.cuhk.edu.hk}
\and
\IEEEauthorblockN{Shaohua Li}
\IEEEauthorblockA{\textit{The Chinese University of Hong Kong}\\ Hong Kong SAR \\ shaohuali@cse.cuhk.edu.hk}
% \and
% \IEEEauthorblockN{4\textsuperscript{th} Given Name Surname}
% \IEEEauthorblockA{\textit{dept. name of organization (of Aff.)} \\
% \textit{name of organization (of Aff.)}\\
% City, Country \\
% email address or ORCID}
% \and
% \IEEEauthorblockN{5\textsuperscript{th} Given Name Surname}
% \IEEEauthorblockA{\textit{dept. name of organization (of Aff.)} \\
% \textit{name of organization (of Aff.)}\\
% City, Country \\
% email address or ORCID}
% \and
% \IEEEauthorblockN{6\textsuperscript{th} Given Name Surname}
% \IEEEauthorblockA{\textit{dept. name of organization (of Aff.)} \\
% \textit{name of organization (of Aff.)}\\
% City, Country \\
% email address or ORCID}
}

\maketitle

\begin{abstract}
Compiler missed optimizations refer to cases in which compilers failed to optimize certain code. 
It takes many compiler developers' efforts to implement or patch such missed optimizations.
In this paper, we present a systematic study of how well agents patch compiler missed optimizations.
We identify a significant challenge that patching a missed optimization requires more than just fixing the reported case, and instead requires generalizing to similar cases.
% We identify a significant challenge: patching a missed optimization requires more than just fixing the reported case; it requires generalizing to similar cases.
% Compiler missed optimization patching requires agents to do more than fix the reported test case: the generated patch should generalize according to developer intent. 
% This paper presents an empirical study of generalization in agent-based compiler missed-optimization patching. 
We construct a benchmark of real-world LLVM missed optimization issues and
compare agent-generated patches with patches from developers in terms of optimization scope.
Our results show that coding agents often optimize the given examples, but many generated patches either cover only part of the developer-intended scope or partially overlap with it; in some cases, they further generalize beyond the reference patch.
% Generic generalization instruction provides limited benefit. 
We further introduce historical-knowledge augmentation techniques that leverage prior LLVM optimization pull requests through retrieval and distillation, showing that they improve developer-aligned generalization and yield practical benefits when applied to real-world IR.

% We further show that historical knowledge from prior LLVM optimization pull requests, provided through retrieval or distillation, improves developer-aligned generalization and increases optimization hits on real-world IR.

% We study this challenge on real-world LLVM missed-optimization issues by comparing agent-generated patches with patches from developers in terms of optimization scope. 

\end{abstract}


\section{Introduction}

Compiler optimization is a fundamental component of modern compiler design and software performance engineering. It aims to improve the efficiency of generated code without altering the intended behavior of the program~\cite{alfred2007compilers}. 
Modern compilers, such as LLVM~\cite{lattner2004llvm}, often perform optimizations by identifying specific code patterns and replacing them with more efficient equivalents.
This process often requires compiler optimization passes to enumerate and match a large number of code patterns in order to identify potential optimization opportunities in the target code.

However, during the design and implementation of compiler optimization rules, many optimization opportunities may remain unrecognized or unexploited by compiler developers, leading to the problem of missed optimizations.
Such missed optimizations can result in suboptimal generated code and have long been a persistent challenge in compiler development~\cite{xu2026lpo}.
To uncover these missed opportunities, researchers and compiler developers have explored a broad spectrum of techniques, including superoptimizers represented by Souper~\cite{sasnauskas2017souper}, synthesis-based systems exemplified by Hydra~\cite{mukherjee2024hydra}, and more recent AI-driven approaches, such as the method proposed by Italiano and Cummins~\cite{italiano2025finding} and LPO~\cite{xu2026lpo}.
As these techniques continue to expose large numbers of missed optimization cases, the main bottleneck is increasingly shifting from identifying optimization opportunities to integrating them as high-quality patches within the compiler maintenance process.

This shift makes automated patching attractive.
In particular, recent advances in agents powered by large language models (LLMs) suggest a promising path toward automatically resolving compiler issues through patch generation, thereby reducing substantial manual engineering effort~\cite{zheng2026agentic}.
These efforts primarily demonstrate the potential of LLM agents in conventional compiler issue resolution, particularly in bug fixing.
However, missed optimization patching is fundamentally different from conventional bug fixing.
For a bug fix, correctness is often approximated by whether the patch satisfies the existing test suite~\cite{weimer2009automatically, smith2015cure, le2016history}.
By comparison, a missed optimization raises a different challenge that the goal is not merely to make the motivating test cases pass, but to implement the optimization in a way that generalizes appropriately beyond those cases. 
For example, a reported missed optimization may expose only a concrete instance of a more general rewrite:
\begin{equation}
\label{eq:generalized-usubsat}
\begin{aligned}
&\mathtt{add}(\mathtt{umax}(a, 10), -10)
\Rightarrow
\mathtt{usub.sat}(a, 10) \\
&\qquad\leadsto\;
\mathtt{add}(\mathtt{umax}(X, C), -C)
\Rightarrow
\mathtt{usub.sat}(X, C).
\end{aligned}
\end{equation}
A less-general patch may match only the concrete instance, including the particular constant \(10\). 
A generalized patch should instead capture the parameterized rule in Eq.~\ref{eq:generalized-usubsat}, which applies across valid choices of \(X\) and \(C\). 
Thus, two patches may both optimize the reported test case while differing in whether they capture the developer-intended optimization scope.

\emph{This difference raises a central challenge: generalization.} In practice, developers often address a missed-optimization issue through a generalization-oriented workflow. They first observe that a particular intermediate representation (IR) fragment admits a more optimized form, while the current compiler fails to perform the corresponding transformation. Starting from this concrete example, developers then infer and generalize the underlying optimization pattern so that it applies to a broader class of related IR cases, rather than only to the single instance initially reported. Finally, they modify the compiler implementation according to the generalized pattern, producing a patch that can resolve this class of optimization opportunities. We therefore define generalization as the process of transforming a specific, ungeneralized optimization into a more general optimization rule that applies not only to the original code fragment but also to similar code scenarios, while preserving the correctness of the optimization. Consequently, an agent-generated patch that passes the given tests may still fail to match developer intent by being too narrow, too broad, or misaligned with the intended optimization pattern.

In this paper, we study whether agents can align their generalization with developer intent when repairing missed-optimization issues via patch generation. Rather than focusing solely on whether agents can generate patches that fix the reported instance, we investigate how task context provided to the agents affects their ability to infer the intended optimization pattern and produce patches with developer-aligned generalization. Guided by this benchmark, we structure our study around the following research questions (RQs):

% To this end, we first construct a benchmark of real-world missed-optimization issues. Each issue in the benchmark is paired with a developer-written patch and relevant test cases, enabling us to assess whether agent-generated patches merely address the reported case or generalize in a way that matches the intent of human developers.
 
\begin{itemize}
    \item \textbf{RQ1:} How well do agents repair real-world missed-optimization issues, and to what extent do their patches generalize in alignment with developer intent?

    \item \textbf{RQ2:} Does generic generalization instruction improve the alignment between agent-generated patches and developer-intended generalization scope?

    \item \textbf{RQ3:} Can historical-knowledge augmentation help agents infer developer-intended generalization scope for missed optimization?

    \item \textbf{RQ4:} Do the alignment improvements enabled by historical knowledge translate into changes in optimization hit behavior on real-world software?
\end{itemize}

Our first two studies show that agents do not reliably align patches with developer intent because they often fail to infer the intended generalization scope, and a generic generalization instruction provides limited benefit and can even worsen misalignment.
These results motivate a further question: \emph{Can historical LLVM optimization experience provide a useful generalization context for the agent?}

% To answer this question, we collect historical LLVM optimization pull requests (PRs) and extract their patches, added regression tests, and summarized generalization behavior. We study two ways of injecting this experience into the agent: retrieval-augmented generation (RAG)~\cite{lewis2020retrieval}, which provides similar source-to-target IR transformations as concrete precedents, and distillation, which summarizes recurring component-level generalization principles as compact background knowledge. Our results show that historical augmentation improves alignment with developer intent across models and changes optimization-hit behavior on real-world IR workloads, suggesting that effective missed-optimization repair requires reusable developer experience rather than only stronger generalization instructions.

To answer this question, we collect historical LLVM optimization pull requests (PRs) and extract their patches, added regression tests, and summarized generalization behavior. We propose two historical-knowledge augmentation techniques that inject such experience into agents: \emph{retrieval-augmented generation (RAG)}~\cite{lewis2020retrieval}, which provides similar source-to-target IR transformations as concrete precedents, and \emph{distillation}, which summarizes recurring component-level generalization principles as compact background knowledge. Our results show that these techniques improve alignment with developer intent across models and yield practical benefits on real-world software, suggesting that effective missed-optimization patching requires reusable developer experience rather than only stronger generalization instructions.

In summary, this paper makes the following contributions:

% \begin{itemize}
%     \item We formulate LLVM missed-optimization repair as a generalization-sensitive patch generation problem, where success requires matching the developer-intended optimization scope.

%     \item We build a benchmark of real-world LLVM missed-optimization issues paired with developer patches and tests for scope-based evaluation.

%     \item We evaluate coding agents under different generalization contexts and show that they frequently misalign with developer intent, while generic generalization instruction provides limited benefit.

%     \item We introduce historical-knowledge augmentation from prior LLVM optimization pull requests and show that it improves intent alignment and affects optimization-hit exposure on real-world IR.
% \end{itemize}
\vspace{-2pt}
\begin{itemize}
    \item We present an empirical study of agent-based LLVM missed-optimization patching; we formulate it as a generalization-sensitive patching problem where success requires matching developer-intended optimization scope.

    \item We construct a benchmark of real-world LLVM missed-optimization issues with patches from developers and regression tests for scope-based evaluation.

    \item We use this benchmark to evaluate coding agents under different generalization contexts, showing that agents frequently misalign with developer intent and that generic generalization instructions provide limited benefit.

    % \item We propose historical-knowledge augmentation techniques from prior LLVM optimization pull requests, based on retrieval and distillation, and show that they improve intent alignment and benefit real-world IR optimization.
    \item We propose historical-knowledge augmentation techniques from prior LLVM optimization PRs via retrieval and distillation, improving intent alignment and real-world IR optimization.
\end{itemize}
\section{Problem Definition}

\subsection{Missed Optimization and Generalization}

A missed optimization occurs when a compiler preserves program semantics but fails to produce a more efficient form. In LLVM, such cases typically appear as issue reports with an initial IR test case \(p_0\). The compiler leaves \(p_0\) suboptimal, while a developer expects it to be simplified by some optimization pass.
Repairing a missed optimization is not merely about making \(p_0\) optimize correctly. The patch must capture the underlying optimization rule and apply it to other related IR whenever the same transformation is valid. We define generalization as the process of inferring a broader optimization pattern from a case, such that the resulting transformation applies not only to the original case but also to semantically similar code patterns, while preserving correctness.

\subsection{Optimization Scope and Assessment}
\label{sec:patch-scope-relationships}

% In this work, we define developer intent as the intended optimization scope that the developer aims to capture when fixing a missed-optimization issue, i.e., the set of programs that the developer expects the patch to optimize while preserving semantics. This latent intent is not directly observable. 
% Therefore, we use the developer-submitted fix as a proxy for this intent and refer to it as the golden patch. We further use the associated tests as representative samples of the intended scope.
% Specifically, we treat the tests added or modified in fixing commit as representative samples of the intended scope, and we use the golden patch's behavior on these tests as a lower-bound indicator of what the developer intended to cover.

In this work, we define developer intent as the intended optimization scope that the developer aims to capture when fixing a missed-optimization issue, i.e., the set of programs that the developer expects the patch to optimize while preserving semantics. 
Since this latent intent is not directly observable, we approximate it using the developer-submitted fix, referred to as the golden patch, and the tests added or modified in the fixing commit, referred to as golden test cases. These golden test cases serve as observable samples of the intended optimization scope and provide a lower-bound approximation.

Each patch induces a partial transformation \(T\) over LLVM IR programs. The optimization scope \(S(T)\) is the set of programs on which \(T\) applies and preserves semantics:
\begin{equation}
    S(T) = \{p \mid T(p) \text{ is defined and } T(p) \equiv p\}.
\end{equation}

For each issue, we compare two patches. The golden patch \(T_G\) is the developer's fix, and its scope \(G = S(T_G)\). The agent patch \(T_A\) is generated by an agent, with scope \(A = S(T_A)\).
\begin{table}[htbp]
\centering
\caption{Observed scope relationships between agent and golden patches based on test-based proxies.}
\small
\begin{tabularx}{\linewidth}{c|X}
\hline
\textbf{Relationship} & \textbf{Interpretation} \\
\hline

\(A \subset G\) & Agent under-generalizes: misses intended cases. \\
\(A \bowtie G\) & Partial overlap: misses some intended cases while adding others outside \(G\). \\
\(G \subset A\) & Agent generalizes beyond the golden scope, potentially covering a broader optimization space. \\
\(A \sim G\) & No observed difference between agent and golden scope.\\

\hline
\end{tabularx}
\label{tab:scope-relations}
\end{table}

Our evaluation asks how the observed behavior of the agent patch relates to the golden patch, i.e., whether the agent captures developer-intended scope as approximated by these proxies, not merely whether it optimizes the initial test case.
In practice, exact enumeration of \(A\) and \(G\) is infeasible. We approximate their relationship using two types of tests, detailed in Section~\ref{sec:generalization-assessment}. Golden tests verify that the agent covers cases the golden patch is expected to optimize, while fuzz tests search for cases where the agent optimizes differently from golden patch, especially cases optimized only by the agent.

We acknowledge that this proxy-based assessment cannot guarantee perfect enumeration of \(A\) and \(G\). A patch passing all golden tests may still miss some intended cases not covered by the tests, and a patch optimizing a fuzz-generated case outside golden tests might represent a valid broader generalization that developers do not explicitly test. Our classification is therefore an empirical characterization of observed equivalence or divergence, not a formal proof of intent.

Based on these signals, we classify the observed relationship between the behavior of the agent patch and the golden patch into four categories, as shown in Table~\ref{tab:scope-relations}. These categories form the basis for our empirical assessment in Section~\ref{sec:generalization-assessment}. 
% We adopt proxy-aware terminology to avoid overclaiming: each category is framed as an interpretation of test results rather than a definitive statement of developer intent.

% \begin{table}[h]
% \centering
% \caption{Observed scope relationships between agent and golden patches (based on test-based proxies).}
% \small
% \begin{tabularx}{\linewidth}{c|X}
% \hline
% \textbf{Relationship} & \textbf{Interpretation} \\
% \hline
% \(A \subset G\) & Agent under-generalizes: misses some golden test cases. \\
% \(A \bowtie G\) & Partial overlap: misses some golden tests but shows unique optimizations on fuzz tests. \\
% \(G \subset A\) & Agent passes all golden tests and shows additional optimizations on fuzz tests (observed broader coverage). \\
% \(A = G\) & Agent passes all golden tests, and no divergence is found via fuzzing (observed equivalent behavior on the evaluated test suite). \\
% \hline
% \end{tabularx}
% \label{tab:scope-relations}
% \end{table}

\section{Study Design}

% \batu{still think this section needs a framework figure...}

To investigate whether agents can infer generalized optimization patterns from initial test cases and generate patches aligned with developer intent, we first construct a benchmark from real-world LLVM missed-optimization issues, then design an agent harness for controlled patch generation, and finally establish a generalization assessment procedure that compares generated patches with golden patches.

\subsection{Benchmark Construction}
\label{sec:benchmark-construction}

We construct our benchmark from real missed-optimization issues in the official LLVM repository~\footnote{\url{https://github.com/llvm/llvm-project/}}. 
Fig.~\ref{fig:bench} schematically illustrates the structure of the LLVM repository on GitHub.
Starting from LLVM GitHub Issues, we collect closed issues labeled both fixed and missed-optimization. We exclude reports that are marked as {duplicate}, {invalid}, or {wontfix}, as well as reports whose fixes target components outside the scope of this study, such as backend-specific code generation, Clang, MLIR, or LLDB. For each remaining issue, we automatically identify the corresponding developer fix from issue discussions, linked pull requests, and commit references. 
In rare cases where this association cannot be inferred reliably from repository metadata, we resolve it manually. 
% When this association cannot be inferred reliably from repository metadata alone, we resolve it manually.

For each selected report, we first record the optimization pass to which the missed optimization belongs. We then extract the developer-written changes from the corresponding fixing commit and use them as the golden patch.
We also collect LLVM IR tests added or modified by fixing commit and use IR input in these tests as source IR of golden test cases. To obtain the corresponding expected IR, we run each source IR through the compiler built after applying the golden patch, using the same optimization pass recorded for the report. The resulting source--expected IR pairs are used as the golden test cases.
In addition, we extract test case from the initial issue report and refer to it as the initial test case, which serves as the missed-optimization example provided to the agent.

We validate each issue before including it in the benchmark. First, we confirm that the pre-fix version exhibits reported missed optimization on the initial test case. Second, after applying the developer fix, we rebuild LLVM and confirm that the regression test passes with the expected optimized output. 

The final benchmark contains 43 verified LLVM missed-optimization instances. Although modest in size, these instances are selected under strict validation criteria, requiring a reproducible initial test, an identifiable developer fix, and recoverable golden tests. This design favors high-confidence scope-based evaluation over a larger but noisier benchmark.
Table~\ref{tab:benchmark-overview} shows their distribution across LLVM optimization passes. InstCombine accounts for the majority of cases because it is LLVM's central pass for local IR canonicalization and peephole simplification, where many missed optimizations are reported and fixed. This is consistent with compiler testing research findings, where InstCombine is the most buggy component~\cite{zhou2021empirical}.
The remaining cases involve SimplifyCFG, ValueTracking, ConstraintElimination, and InstructionSimplify.
\vspace{-5pt}

\begin{table}[htbp]
    \centering
    \caption{Benchmark instances by LLVM Passes.}
    \label{tab:benchmark-overview}
    \small
    \setlength{\tabcolsep}{4pt}
    \begin{tabular}{lclc}
        \toprule
        \textbf{Component} & \textbf{\#} &
        \textbf{Component} & \textbf{\#} \\
        \midrule
        InstCombine & 32 &
        SimplifyCFG & 4 \\
        ValueTracking & 3 &
        ConstraintElimination & 2 \\
        InstructionSimplify & 2 &
        \multicolumn{2}{c}{} \\
        \bottomrule
    \end{tabular}
\vspace{-10pt}
\end{table}

\begin{figure*}[htbp]
    \noindent
    \begin{minipage}[t]{0.66\linewidth}
        \centering
        \vspace{0pt}
        \includegraphics[width=\linewidth]{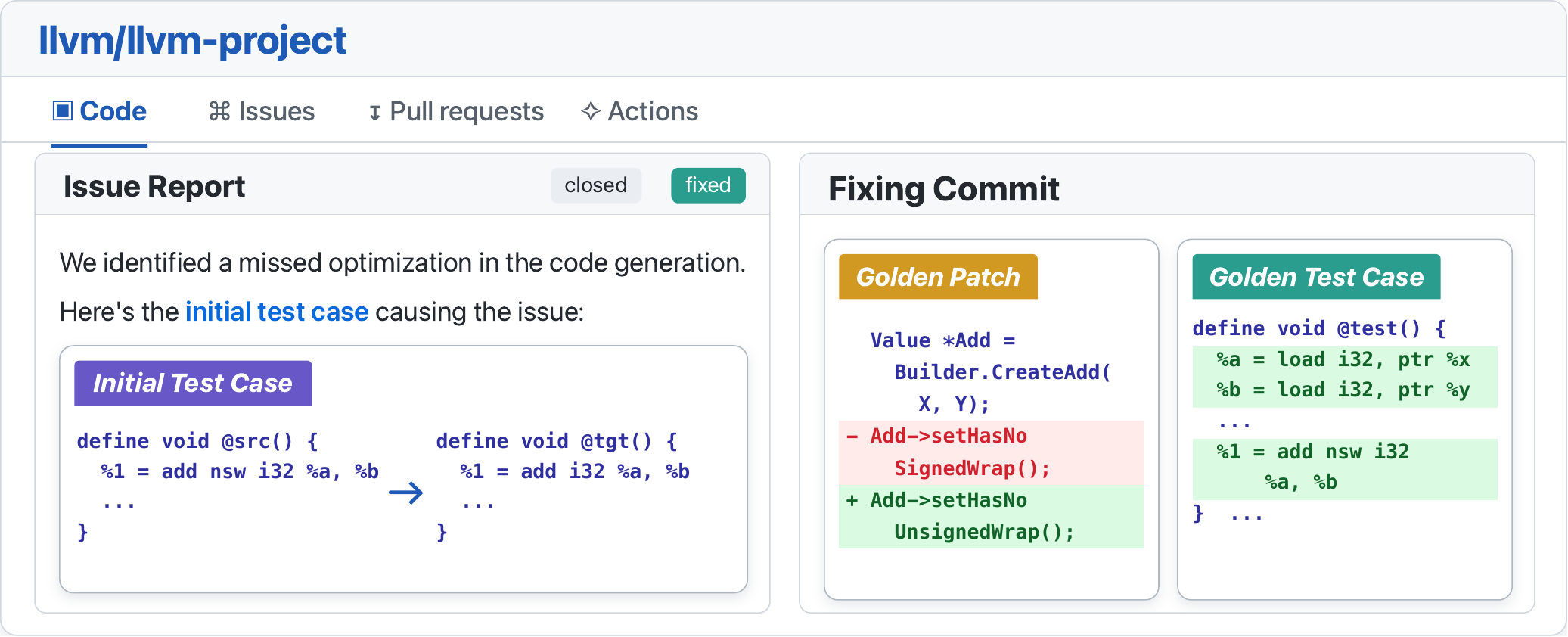}
        \caption{Illustration of how LLVM repository corresponds to benchmark components.}
        \label{fig:bench}
    \end{minipage}
    \hfill
    \begin{minipage}[t]{0.30\linewidth}
        \centering
        \vspace{0pt}

        % Smaller value raises the table block.
        % Try 0.10\linewidth, 0.05\linewidth, or 0pt.
        \vspace{-0.5em}

        \captionof{table}{Tool categories exposed by our agent harness.}
        \label{tab:baseline-agent-tools}
        \vspace{0.3em}

        \scriptsize
        \setlength{\tabcolsep}{2pt}
        \renewcommand{\arraystretch}{1.05}
   \begin{tabular}{@{}
    >{\raggedright\arraybackslash}p{0.22\linewidth}
    >{\raggedright\arraybackslash}p{0.33\linewidth}
    >{\raggedright\arraybackslash}p{0.37\linewidth}
@{}}
    \toprule
    \textbf{Category} & \textbf{Tool} & \textbf{Purpose} \\
    \midrule
    Context &
    \texttt{issue\_info}, \texttt{view\_source}, \texttt{get\_langref} &
    Inspect tasks, source code and LLVM docs. \\
    \midrule
    Editing &
    \texttt{apply\_code} &
    Apply localized edits to LLVM source files. \\
    \midrule
    Validation &
    \texttt{build\_and} \texttt{\_check}, \texttt{lit\_check} &
    Build, test, and validate candidate optimizations. \\
    \midrule
    Workflow &
    \texttt{show\_diff}, \texttt{submit\_patch} &
    Review and submit generated patches. \\
    \bottomrule
\end{tabular}
    \end{minipage}
    \vspace{-15pt}
\end{figure*}

\subsection{Agent Harness}
\label{sec:agent-harness}

For each issue, we would like to evaluate an agent’s ability to correctly implement the missed optimization. To provide a simple and reproducible reference point, we implement a lightweight agent harness for LLVM missed-optimization issues. The harness connects a model to an issue-specific LLVM workspace, a controlled tool interface, and a validation loop. If the agent exits before producing a valid patch, exceeds its interaction budget, or fails final validation, the run is counted as a failure. Otherwise, the validated diff is recorded as an accepted patch.

\subsubsection{Environment Setup}
For a benchmark instance, the harness prepares an LLVM workspace at the parent commit of the developer fix. To facilitate the agent, we also implement a set of tools, as Table~\ref{tab:baseline-agent-tools} lists, which the agent can use to interact and modify compilers. The agent is given access to issue-specific context, relevant LLVM source files, and LLVM Language Reference Manual~\cite{llvm-langref}, but not to the golden patch or golden test cases. 

\subsubsection{Overall Workflow}
% After environment setup, the agent is instructed to act as an LLVM developer and give the reported missed optimization a patch. Starting from the initial test case, it first infers the likely optimization pass, such as InstCombine, SimplifyCFG, or ValueTracking, and receives lightweight localization hints in the form of ranked candidate files and functions. These hints narrow the initial search space because missed-optimization fixes are often concentrated in the responsible optimization pass or analysis,\footnote{For example, fixes for InstCombine optimizations are often located under \texttt{llvm/lib/Transforms/InstCombine}.} but the agent may inspect other files when necessary. 

After environment setup, the agent is instructed to act as an LLVM developer and generate a patch for the reported missed optimization. 
% Starting from the initial test case, it first infers the likely optimization pass, such as InstCombine, SimplifyCFG, or ValueTracking, and then locates the corresponding candidate files and functions based on that pass. 
% When the agent begins generating a patch, it starts analyzing the necessary modifications at the locations of the candidate functions.
% Since missed-optimization fixes are often concentrated in the responsible optimization pass or analysis~\footnote{For example, fixes for InstCombine optimizations are often located under \texttt{llvm/lib/Transforms/InstCombine}.}, this narrows the initial search space, but the agent may inspect other files when necessary.
Starting from the initial test case, it first infers the optimization pass and the functions to patch.
The agent then enters a ReAct loop~\cite{yao2022react} that interleaves root-cause analysis, source inspection, patch synthesis, and validation. It uses \texttt{issue\_info} to inspect benchmark instance, \texttt{view\_source} to examine LLVM source files, and \texttt{get\_langref} when the optimization depends on LLVM IR semantics. After identifying a plausible missing optimization rule, the agent applies an edit with \texttt{apply\_code} and invokes \texttt{build\_and\_check} to validate candidate patch. Validation checks correctness with Alive2~\cite{lopes2021alive2} and estimates effectiveness with llvm-mca~\cite{llvm_mca_doc} using block-throughput cost. A patch passes only if its optimized output is no more costly than either original output or golden optimized output. We also run lit~\cite{llvm-lit-doc} to detect regressions, but treat lit failures as soft feedback because FileCheck patterns can be sensitive to non-semantic textual differences. Build errors, lit failures, Alive2 failures, and llvm-mca regressions are returned to the agent for revision. The agent may inspect changes with \texttt{show\_diff} and submit the final patch with \texttt{submit\_patch}; the loop ends upon submission or budget exhaustion.

\subsection{Generalization Assessment}

\label{sec:generalization-assessment}
% We assess the generalization of an agent-generated patch along two dimensions: whether it covers the developer-intended optimization pattern and whether it diverges beyond the scope of the golden patch.

We assess the generalization of an agent-generated patch by characterizing its scope relationship with the golden patch, as defined in Section~\ref{sec:patch-scope-relationships}. Given an agent-generated patch \(A\) and a human-developed golden patch \(G\), our goal is to determine whether \(A\) has the same optimization scope as \(G\), is less general than \(G\), or is more general than \(G\). We accomplish this evaluation using two complementary sources of tests, including golden tests and newly generated fuzz tests.

First, we run golden test cases extracted during benchmark construction on agent-patched compiler.
Since these test cases are validated by golden patch, we regard them as representative of the optimization pattern that the golden patch is intended to capture.
Therefore, golden test cases are withheld during agent patch generation and used only for post-hoc evaluation, so that they do not bias or interfere with the agent's generalization decisions.
If an agent-generated patch fails to optimize these cases, we consider it to exhibit insufficient generalization in the optimization direction intended by the developer.

% Second, we fuzz agent-patched compiler to identify potential divergent or excessive generalization, as illustrated in Figure~\ref{fig:fuzz}.
% We use two complementary fuzzing strategies.
% The traditional fuzzer starts from the initial test case as the seed IR and applies randomized mutations to generate new LLVM IR programs.
% In contrast, the LLM-based generator takes the agent-generated patch and the golden patch as input; it is prompted to compare the two patches and synthesize an LLVM IR program that may expose a behavioral difference between them.
% Each generated program is then optimized by both the golden-patched and agent-patched LLVM compilers.
% For each candidate program, we first run Alive2 to verify that the optimized IR is a semantically valid refinement of the source, and then check the efficiency of the optimized IR using llvm-mca.
% If the agent patched compiler yields a lower llvm-mca computed cost for a program than the golden patched compiler, we treat it as evidence that the agent-generated patch optimizes cases outside the developer-intended scope.

\begin{figure}[htbp]
    \centering
    \includegraphics[width=\linewidth]{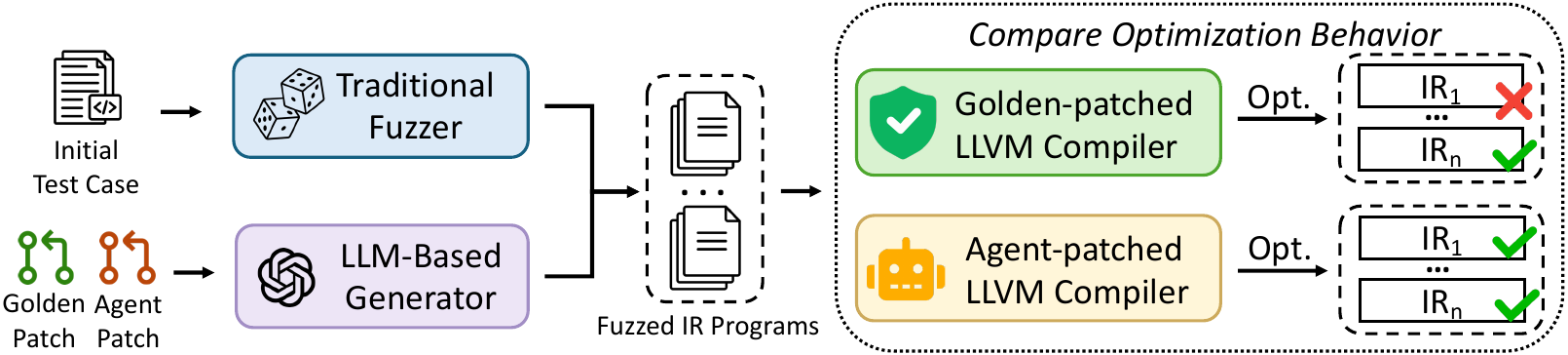}
    \caption{Fuzz-based generalization assessment. $\text{IR}_1$ will be identified as a program outside the developer-intended scope.}
    \label{fig:fuzz}
  \vspace{-5pt}
\end{figure}

Second, we fuzz the agent-patched compiler to identify potential divergent or excessive generalization, as illustrated in Figure~\ref{fig:fuzz}.
We use two complementary fuzzing strategies.
The traditional fuzzer starts from the initial test case as the seed IR and applies randomized mutations to generate new LLVM IR programs.
Inspired by WhiteFox~\cite{yang2024whitefox}, we further design an LLM-based generator to fuzz LLVM IR programs that are likely to expose behavioral differences between the agent-generated patch and the golden patch.
Specifically, the LLM-based generator takes the agent-generated patch and the golden patch as input; it is prompted to compare the two patches and synthesize an LLVM IR program that may expose a behavioral difference between them.
Each generated program is then optimized by both the golden-patched and agent-patched LLVM compilers.
For each candidate program, we first run Alive2 to verify that the optimized IR is a semantically valid refinement of the source, and then check the efficiency of the optimized IR using llvm-mca.
If the agent-patched compiler yields a lower llvm-mca computed cost for a program than the golden-patched compiler, we treat it as evidence that the agent-generated patch optimizes cases outside the developer-intended scope.

We use these two signals to classify generalization relationship between the agent patch scope \(A\) and the golden patch scope \(G\), following Section~\ref{sec:patch-scope-relationships}. Specifically, failing the golden test cases with no agent-only fuzz cases suggests \(A \subset G\); failing the golden test cases with agent-only fuzz cases suggests a proper intersection between \(A\) and \(G\); passing the golden test cases with agent-only fuzz cases suggests \(G \subset A\); and passing the golden test cases with no agent-only fuzz cases suggests likely \(A \sim G\), which denotes no observed behavioral difference under our evaluation.

\section{Baseline Study}

\subsection{Research Questions}
\label{sec:research-questions}

We study whether agents can generate patches whose optimization behavior aligns with the developer intent embodied by corresponding golden patch.
% We consider two settings: a direct setting in which agents receive no explicit generalization context, and a context-enhanced setting in which agents receive high-level guidance asking for generalization.
We consider two settings. In the direct setting, agents receive no explicit generalization context. In the context-enhanced setting, agents receive high-level guidance asking for generalization.

\begin{itemize}
   \item \textbf{RQ1:} How well do agents repair real-world missed-optimization issues, and to what extent do their patches generalize in alignment with developer intent?

    \item \textbf{RQ2:} Does generic generalization instruction improve the alignment between agent-generated patches and developer-intended generalization scope?
\end{itemize}

\subsection{Experimental Setup}

We evaluate all agent settings under a controlled setup that fixes the benchmark, tool interface, validation feedback, and interaction budget. This setup isolates the effects of the base model and additional instructions.

\subsubsection{Models and Agent Framework}
We select a variety of both closed and open-source LLMs. Specifically, we evaluate agent performance with four models: GPT-5.5~\cite{openai2026gpt55systemcard}, DeepSeek-V4-Pro~\cite{deepseek2026deepseek}, Qwen3.5-Plus~\cite{qwen35blog} and Kimi K2.5~\cite{team2026kimi}. All models use the same harness, issue context, localization hints, tools, and feedback. The agent and the localizer always employ the same model.
We use mini-SWE-agent~\cite{yang2024sweagent} as the coding-agent framework in our experiments. 
% For each report, the agent receives the issue description and initial test case, interacts with LLVM through fixed tools, and iteratively revises a patch.

\subsubsection{Fuzzing Settings}
For the fuzzing component, we use alive-mutate~\cite{fan2024high} as the traditional fuzzer. Given an input seed program, we generate 1,000 mutants. For the LLM-based generator, considering cost constraints, we use Qwen3.5-Plus. 
For each patch pair, we ask the model to generate four test cases and repeat this 15 times, totaling 60 test cases.

\subsubsection{Run Configuration}
For each model and each benchmark instance, we run the baseline agent with a fixed interaction budget of 50 steps. Each step corresponds to one ReAct iteration~\cite{yao2022react}, in which the model proposes one or more bash commands and receives the resulting tool output as feedback. A run terminates when the agent explicitly signals submission, exhausts the 50-step budget, exceeds the configured cost limit, or encounters a runtime error.
All llvm-mca evaluations are performed under the x86\_64 Skylake configuration.

% Build or test failures during the
% repair loop do not terminate the run by themselves; instead,
% the harness returns the corresponding feedback to the agent for
% further revision. Runs that end without an explicit submission
% are recorded as incomplete repair attempts.

\subsubsection{Agent Instructions}
\label{sec:prompt-conditions}

We evaluate agents under two instruction conditions that differ only in the task-level guidance provided to the agent.
In the default condition, the agent is instructed to act as an LLVM developer, diagnose the reported missed optimization, inspect the relevant LLVM source code, implement a minimal fix, and validate the patch using the provided build and test tools.
The instructions also specify the available tool interface, the expected edit-test workflow, and constraints such as modifying only LLVM implementation and interface files within the benchmark scope.

To study whether explicit instruction-based generalization context improves alignment with developer-intended optimization scope (RQ2), we add 
% one sentence
explicit guidance
to default instructions: \emph{``Ensure that the final generated patch can handle more general patterns, rather than being limited to the provided test cases.''}
We intentionally avoid specifying the developer intent in instructions, because our goal is to evaluate whether the agent can, after being encouraged to generalize, infer and align with the intended optimization scope by itself from the issue context and codebase.
Apart from this added instruction, the task description, tool interface, validation procedure, and interaction budget remain unchanged.

\subsection{Empirical Findings}

We now present the empirical results for the two research questions introduced in Section~\ref{sec:research-questions}. We first examine patch generalization under the default context, and then evaluate the effect of generalization-oriented instructions.

\subsubsection{RQ1. Generalization Behavior under Default Context}
We first evaluate the performance of agents without providing any additional instructions or guidance related to generalization. Table~\ref{tab:rq1} reports the performance of agents across different base models. 
The table reports, for each base model, the number of issues whose agent-generated patch exhibits each relationship with the corresponding golden patch, following the definitions in Section~\ref{sec:patch-scope-relationships}. The Failed column denotes the number of issues for which the agent failed to generate, within the given budget, a patch that optimizes the initial test case.

% \begin{table}[htbp]
%     \centering
%     \caption{Issue counts by patch-scope relationship across base models under direct patch generalization.}
%     \label{tab:rq1}
%     \small
%     \resizebox{\columnwidth}{!}{
%     \begin{tabular}{lccccc}
%         \toprule
%         \textbf{Model} &
%         \(\mathbf{A \subset G}\) &
%         \(\mathbf{A \bowtie G}\) &
%         \(\mathbf{G \subset A}\) &
%         \(\mathbf{A = G}\) &
%         \textbf{Failed} \\
%         \midrule
%         GPT-5.5 & 4 & 4 & 6 & 18 & 11 \\
%         DeepSeek-V4-Pro & 5 & 4 & 4 & 14 & 16 \\
%         Qwen3.5-Plus & 9 & 4 & 3 & 11 & 16 \\
%         Kimi K2.5 & 11 & 3 & 2 & 10 & 17 \\
%         \bottomrule
%     \end{tabular}
%     }
%     \vspace{-5pt}
% \end{table}

% \begin{table}[t]
%     \centering
%     \caption{Issue counts by patch-scope relationship across base models under the default instruction setting.}
%     \label{tab:rq1}
%     \small
%     \resizebox{\columnwidth}{!}{
%     \begin{tabular}{lcccccc}
%         \toprule
%         \textbf{Model} &
%         \(\mathbf{A \subset G}\) &
%         \(\mathbf{A \bowtie G}\) &
%         \(\mathbf{G \subset A}\) &
%         \(\mathbf{A \sim G}\) &
%         \textbf{Total} &
%         \textbf{Failed} \\
%         \midrule
%         GPT-5.5 & 4 & 4 & 6 & 18 & 32 & 11 \\
%         DeepSeek-V4-Pro & 5 & 4 & 4 & 14 & 27 & 16 \\
%         Qwen3.5-Plus & 9 & 4 & 3 & 11 & 27 & 16 \\
%         Kimi K2.5 & 11 & 3 & 2 & 10 & 26 & 17 \\
%         \bottomrule
%     \end{tabular}
%     }
% \end{table}

\begin{table}[htbp]
    \centering
    \caption{Issue counts by patch-scope relationship across base models under the default instruction setting.}
    \label{tab:rq1}
    \small
    \resizebox{\columnwidth}{!}{
    \begin{tabular}{lcccccc}
        \toprule
        \multirow{2}{*}{\textbf{Model}} &
        \multicolumn{5}{c}{\textbf{Applied}} &
        \multirow{2}{*}{\textbf{Failed}} \\
        \cmidrule(lr){2-6}
        & \(\mathbf{A \subset G}\) &
        \(\mathbf{A \bowtie G}\) &
        \(\mathbf{G \subset A}\) &
        \(\mathbf{A \sim G}\) &
        \textbf{Total} & \\
        \midrule
        GPT-5.5 & 4 & 4 & 6 & 18 & 32 & 11 \\
        DeepSeek-V4-Pro & 5 & 4 & 4 & 14 & 27 & 16 \\
        Qwen3.5-Plus & 9 & 4 & 3 & 11 & 27 & 16 \\
        Kimi K2.5 & 11 & 3 & 2 & 10 & 26 & 17 \\
        \bottomrule
    \end{tabular}
    }
    \vspace{-5pt}
\end{table}

From Table~\ref{tab:rq1}, we make the following observations. First, in most cases, the agents can successfully generate a patch that can be applied to LLVM and can optimize the initial test case. Specifically, when paired with GPT-5.5, the agent generates such a patch for 74\% (32/43) of the issues. This result demonstrates the agents' strong capability in understanding the project context and producing effective patches for the initial test cases.
However, among all patches that successfully pass the checks based on the initial test cases, only about half fall into \(A \sim G\), where the agent patch matches the optimization scope of the corresponding golden patch. 
A small number of patches fall into \(G \subset A\), as they cover the golden scope and additionally optimize cases beyond it, suggesting potentially broader valid generalization.
The remaining patches exhibit scope mismatches. In \(A \subset G\), the agent patch covers only part of the golden scope, indicating under-generalization. In \(A \bowtie G\), the two patches partially overlap, with the agent patch captures some cases covered by the golden patch but misses others or includes different cases outside the golden scope.

\findingbox{Finding 1}{Agents often generate patches that optimize the initial test case, yet they do not necessarily match the golden patch's optimization scope.}

\subsubsection{RQ2. Generic Generalization Instruction }

In RQ1, we observed that agent-generated patches do not always match the optimization scope of the corresponding golden patches, even when they optimize the initial test cases. However, under the default instruction setting, the agent is not explicitly instructed to consider generalization; it is only asked to modify the LLVM source code so that the specified test case can be optimized. 
Therefore, in this RQ, we investigate whether adding a generic and explicit generalization instruction improves optimization-scope alignment with the golden patch.

After incorporating the generalization instruction described in Section~\ref{sec:prompt-conditions}, we run the agents again and re-evaluate the generated patches relative to the corresponding golden patches. Table~\ref{tab:rq2} reports the resulting issue counts by patch-scope relationship across base models.
\begin{table}[htbp]
    \centering
    \caption{Issue counts by patch-scope relationship under default and generalization instructions. Gen. Inst. is short for Generalization Instruction. ``No'' rows are copied from Table~\ref{tab:rq1} for easy presentation.}
    \label{tab:rq2}
    \small
    \resizebox{\columnwidth}{!}{
    \begin{tabular}{lcccccc}
        \toprule
        \textbf{Model} &
        \textbf{Gen. Inst.} &
        \(\mathbf{A \subset G}\) &
        \(\mathbf{A \bowtie G}\) &
        \(\mathbf{G \subset A}\) &
        \(\mathbf{A \sim G}\) &
        \textbf{Failed} \\
        \midrule
        \multirow{2}{*}{GPT-5.5}
            & No  & 4 & 4 & 6 & 18 & 11 \\
            & Yes & 4 & 2 & 7 & 17 & 13 \\
        \midrule
        \multirow{2}{*}{DeepSeek-V4-Pro}
            & No  & 5 & 4 & 4 & 14 & 16 \\
            & Yes & 8 & 6 & 5 & 10 & 14 \\
        \midrule
        \multirow{2}{*}{Qwen3.5-Plus}
            & No  & 9  & 4 & 3 & 11 & 16 \\
            & Yes & 10 & 6 & 2 & 8  & 17 \\
        \midrule
        \multirow{2}{*}{Kimi K2.5}
            & No  & 11 & 3 & 2 & 10 & 17 \\
            & Yes & 5  & 4 & 1 & 9  & 24 \\
        \bottomrule
    \end{tabular}
    }
    % \vspace{-15pt}
    
\end{table}
% \begin{figure}[htbp]
%   \centering
%   % 第一行
%   \begin{subfigure}[b]{0.48\linewidth}
%     \includegraphics[width=\linewidth]{fig/rq2_GPT-5.5.pdf}
%     \caption{GPT-5.5}
%     \label{fig:rq2-gpt}
%   \end{subfigure}\hfill
%   \begin{subfigure}[b]{0.48\linewidth}
%     \includegraphics[width=\linewidth]{fig/rq2_DeepSeek.pdf}
%     \caption{DeepSeek-V4-Pro}
%     \label{fig:rq2-deepseek}
%   \end{subfigure}

%   \medskip

%   % 第二行
  
%   \begin{subfigure}[b]{0.48\linewidth}
%     \includegraphics[width=\linewidth]{fig/rq2_Qwen3.5-Plus.pdf}
%     \caption{Qwen3.5-Plus}
%     \label{fig:rq2-qwen}
%   \end{subfigure}\hfill
%   \begin{subfigure}[b]{0.48\linewidth}
%     \includegraphics[width=\linewidth]{fig/rq2_Kimi_K2.5.pdf}
%     \caption{Kimi K2.5}
%     \label{fig:rq2-kimi}
%   \end{subfigure}

%   \caption{xxxxxxxOutcome transitions after generalization-oriented prompting.
%     Each panel uses the union of issues from the w/o prompt and w/ prompt runs.
%     Issues appearing in only one run are connected to \textit{Failed}.}
%   \label{fig:rq2}
% \end{figure}

As shown in Table~\ref{tab:rq2}, adding a generalization instruction does not improve coverage of the developer-intended scope. Because \(G \subset A\) is treated as a broader generalization, we consider patches that at least cover the golden scope, i.e., those falling into either \(A \sim G\) or \(G \subset A\). Surprisingly, the number of such patches does not increase for any base model and even decreases for 3 of the 4 models.

Meanwhile, the number of issues where the agent-generated patch has a narrower optimization scope than the golden patch, \(A \subset G\), increases for DeepSeek-V4-Pro and Qwen3.5-Plus. For GPT-5.5 and Kimi K2.5, although \(A \subset G\) decreases, the generalization instruction substantially increases the number of Failed issues, indicating that many generated patches no longer optimize the initial test cases derived from the issue descriptions. These results suggest that explicitly instructing the agent to generalize does not reliably improve coverage of the developer-intended optimization scope. In some cases, it may instead shift failures from scope mismatch to initial-test failure or lead to more under-generalized patches.

\findingbox{Finding 2}{Generic generalization instructions do not reliably improve golden-patch scope alignment and may reduce success of initial test cases.}

\section{Historical-Knowledge-Guided Generalization}

In the previous sections, we showed that agents do not consistently generate patches that both optimize the initial test case and match the developer-intended optimization scope. Although agents can often address the concrete optimization example described in an issue report, they frequently fail to infer the broader optimization scope captured by the corresponding golden patch. Moreover, adding a generalization-oriented instruction does not reliably improve scope alignment and can even degrade performance in some cases.

One possible explanation is that the agents lack access to the knowledge that compiler developers implicitly rely on when implementing optimization rules. In practice, compiler developers rarely design optimizations in isolation. Instead, they build upon accumulated experience from previously implemented transformations, existing optimization patterns, and historical fixes within the codebase. Such historical knowledge often provides valuable clues about how a newly discovered optimization opportunity should be generalized.

Motivated by this observation, we investigate whether providing historical knowledge can help agents better align their generated patches with developer intent. Specifically, we augment the agent with optimization-related experience collected from prior developer fixes and evaluate whether this additional knowledge improves the agent's ability to generalize beyond the issue-level examples. Through this study, we aim to answer the following research question:

\begin{itemize}
 \item \textbf{RQ3:} Can historical-knowledge augmentation help agents infer developer-intended generalization scope for missed optimization?
\end{itemize}

% To answer this question, we adopt two complementary methods to extract and provide information from developers’ historical experience to the agent: RAG and historical knowledge distillation. The former supplies the agent with relevant prior optimization patches retrieved from the codebase history, while the latter distills recurring optimization patterns and generalization principles from historical developer changes into a more compact form. In the following sections, we describe the sources of this historical information and explain how each form of knowledge is constructed and incorporated into the agent’s patch-generation process.

% To answer this question, we adopt two complementary methods to extract information from developers' historical experience and inject it into the agent: RAG and historical knowledge distillation. The former, based on the currently most mainstream retrieval-augmented paradigm~\cite{lewis2020retrieval}, directly retrieves and provides relevant historical optimization patches to the agent as example references; the latter, following the recent trend of skill design~\cite{wang2023voyager}, condenses generalized patterns that repeatedly appear in historical changes into compact high-level rules. Together, they provide historical guidance at the levels of concrete cases and general principles, respectively. In the following sections, we will describe the sources of this historical information and explain in detail how each form of knowledge is constructed and integrated into the agent’s patch generation process.

To answer this question, we adopt two complementary historical-knowledge augmentation methods: \emph{\textbf{RAG}} and \emph{\textbf{distillation}}. \emph{\textbf{RAG}}, based on the retrieval-augmented paradigm~\cite{lewis2020retrieval}, provides relevant historical optimization patches as concrete examples, while \emph{\textbf{distillation}}, following the recent trend of skill design~\cite{wang2023voyager}, condenses recurring patterns in historical changes into compact high-level rules. Together, they provide case-level and principle-level guidance, respectively. Although many other historical-experience augmentation designs are possible, exhaustively evaluating them is beyond our scope. We therefore focus on these two representative and complementary techniques and next describe how their historical knowledge is constructed and integrated into the agent's patch generation process.

\subsection{Historical-Knowledge Augmentation}

\subsubsection{Historical Knowledge Source}
\label{sec:historical-knowledge-source}
% We extract historical instances using a procedure that is largely consistent with the benchmark-construction process described in Section~\ref{sec:benchmark-construction}. The key difference is that, in this study, we collect instances from GitHub pull requests rather than from issues. This design choice is motivated by the need to obtain a substantially larger pool of historical optimization examples. While issue reports often provide clearer problem descriptions and more explicitly structured examples, they are relatively limited in number. Pull requests, in contrast, record a broader range of developer-implemented optimization changes and therefore provide a richer source of historical experience. In total, we collect 869 pull requests as the source of historical optimization instances. 

We extract historical instances using a procedure largely consistent with the benchmark-construction process described in Section~\ref{sec:benchmark-construction}. Unlike benchmark construction, this study collects instances from GitHub pull requests rather than issues, as pull requests offer a larger pool of developer-implemented optimization changes. We exclude all pull requests that are used in benchmark to avoid data leakage. In total, we collect 869 pull requests as the source of historical optimization instances.

For each pull request, we collect source-code patch and accompanying tests, and summarize optimization’s generalization behavior for retrieval. Given the patch and tests, the LLM identifies the key pattern, including the initial concrete case, the additional cases covered by the patch of PR, and the transformation logic connecting them. Since pull requests do not explicitly distinguish the pre-generalization test from additional golden tests, we use a simple heuristic under which the first test case is treated as the initial case, while the remaining tests are treated as additional golden cases. Although this heuristic may not exactly match the developer’s test-writing order, it consistently approximates how the patch generalizes from the original instance to broader coverage.

Overall, each historical instance consists of the source-code patch, the extracted test cases, and an LLM-generated summary of the optimization's generalization details. These instances constitute the historical knowledge source used by both RAG-based augmentation and historical knowledge distillation in the subsequent sections.

\subsubsection{RAG-Based Augmentation}
\label{sec:rag-based}

\begin{figure}[t]
    \centering
    \includegraphics[width=\linewidth]{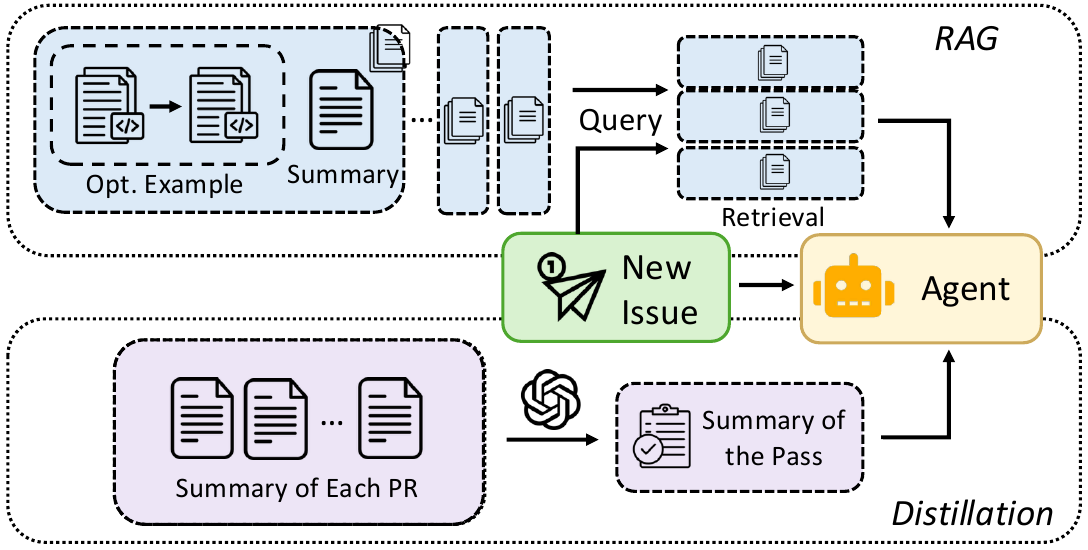}
    \caption{Workflow of RAG- and distillation-based augmentation.}
    \label{fig:history}
    \vspace{-20pt}
\end{figure}

We use RAG to expose agent to relevant historical optimization experience. 
The intuition is that RAG provides agent with relevant project-specific examples at patch-generation time, helping it infer intended  scope from similar historical patches and test cases.
Fig.~\ref{fig:history} shows the workflow of RAG-based augmentation.
For each historical pull request, we index the optimization behavior reflected by its regression tests. Specifically, each test is represented as a structured source-target IR pair, where the source side denotes the original IR pattern and the target side denotes the expected optimized form. To avoid retrieving examples based on testing syntax rather than optimization semantics, we remove FileCheck directives from the target-side test body before indexing. The resulting source-target pair is serialized using the same representation for both historical documents and online queries, ensuring that retrieval is driven by comparable optimization behavior.

% We embed these serialized IR pairs using a embedding model and normalize all vectors to enable cosine-similarity search. At query time, the initial IR test case from the new issue is encoded in the same representation, and the retriever returns the top-\(k\) most similar historical examples. When multiple retrieved examples come from the same pull request, we optionally keep only the highest-scoring one to improve diversity.

We embed all historical examples under this representation and normalize the resulting vectors for cosine-similarity search. Given a new issue, we encode its initial IR test case in the same format and retrieve the top-\(k\) nearest historical examples. To improve diversity, we optionally keep only the highest-scoring example from each pull request.

For each retrieved example, we attach the corresponding historical summary. Thus, the agent receives both concrete behavioral evidence, i.e., the source-to-target IR transformation, and a higher level description of the developer's generalization intent. These retrieved examples are provided as contextual references during patch generation, encouraging the agent to reason by analogy with prior developer-approved optimizations rather than relying solely on the issue-level test case.

% Algorithm~\ref{alg:rag} summarizes the RAG-based augmentation procedure. 
% This design deliberately retrieves examples by optimization behavior rather than by source-code textual similarity. Since two missed-optimization fixes may implement analogous transformations in different LLVM components or through different code paths, IR-level source-to-target similarity provides a more direct signal for identifying relevant historical generalization knowledge.

% \begin{algorithm}[t]
% \caption{RAG-based Historical Knowledge Retrieval}
% \label{alg:rag}
% \begin{algorithmic}[1]
% \REQUIRE Historical records \(\mathcal{H}\), query IR pair \((s,t)\), top-\(k\)
% \ENSURE Retrieved historical examples \(\mathcal{R}\)
% \STATE \(\mathcal{D} \leftarrow \emptyset\)
% \FORALL{\(h \in \mathcal{H}\)}
%     \FORALL{\((s_h,t_h) \in \textsc{Tests}(h)\)}
%         \STATE \(t_h \leftarrow \textsc{StripFileCheck}(t_h)\)
%         \STATE \(d_h \leftarrow \textsc{SerializeIRPair}(s_h,t_h)\)
%         \STATE \(\mathcal{D} \leftarrow \mathcal{D} \cup \{(d_h,h)\}\)
%     \ENDFOR
% \ENDFOR
% \STATE \(E \leftarrow \textsc{Normalize}(\textsc{Encode}(\mathcal{D}))\)
% \STATE \(q \leftarrow \textsc{SerializeIRPair}(s,t)\)
% \STATE \(e_q \leftarrow \textsc{Normalize}(\textsc{EncodeQuery}(q))\)
% \STATE \(\mathcal{R} \leftarrow \textsc{TopKByCosine}(\mathcal{D},E,e_q,k)\)
% \IF{deduplication is enabled}
%     \STATE \(\mathcal{R} \leftarrow \textsc{DeduplicateByPR}(\mathcal{R})\)
% \ENDIF
% \STATE \(\mathcal{R} \leftarrow \textsc{AttachSummaries}(\mathcal{R})\)
% \RETURN \(\mathcal{R}\)
% \end{algorithmic}
% \end{algorithm}

\begin{figure*}[htbp]
  \centering
  \captionsetup[subfigure]{skip=2pt}
  \captionsetup{skip=4pt}

  % 第一行：Baseline vs. RAG
  \begin{subfigure}[b]{0.24\textwidth}
    \includegraphics[width=\linewidth]{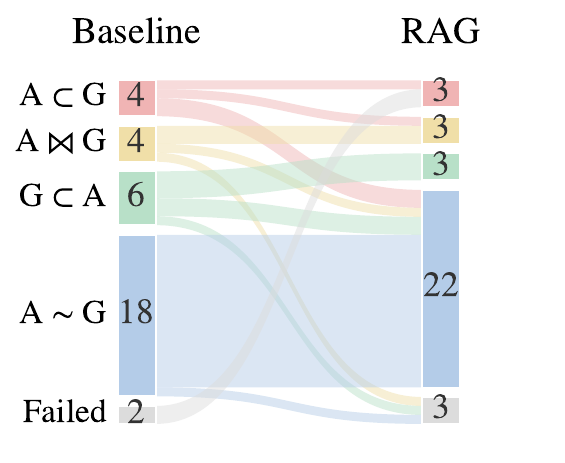}
    \caption{GPT-5.5}
    \label{fig:rq3-gpt-rag}
  \end{subfigure}\hfill
  \begin{subfigure}[b]{0.24\textwidth}
    \includegraphics[width=\linewidth]{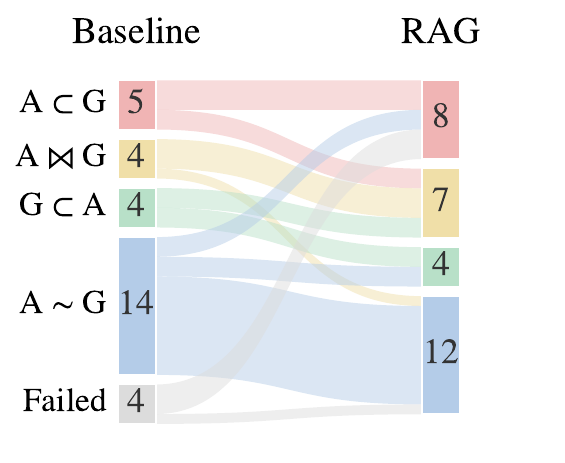}
    \caption{DeepSeek-V4-Pro}
    \label{fig:rq3-deepseek-rag}
  \end{subfigure}\hfill
  \begin{subfigure}[b]{0.24\textwidth}
    \includegraphics[width=\linewidth]{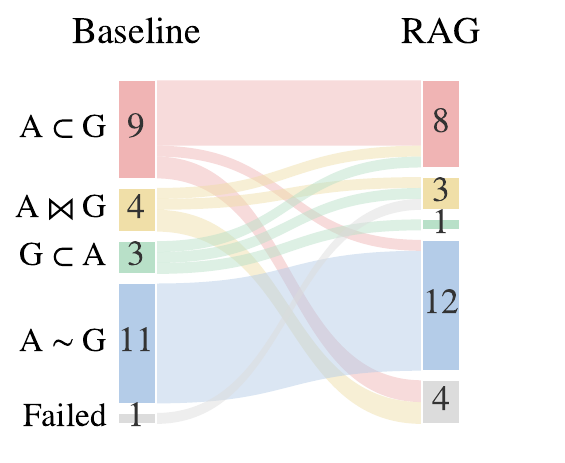}
    \caption{Qwen3.5-Plus}
    \label{fig:rq3-qwen-rag}
  \end{subfigure}\hfill
  \begin{subfigure}[b]{0.24\textwidth}
    \includegraphics[width=\linewidth]{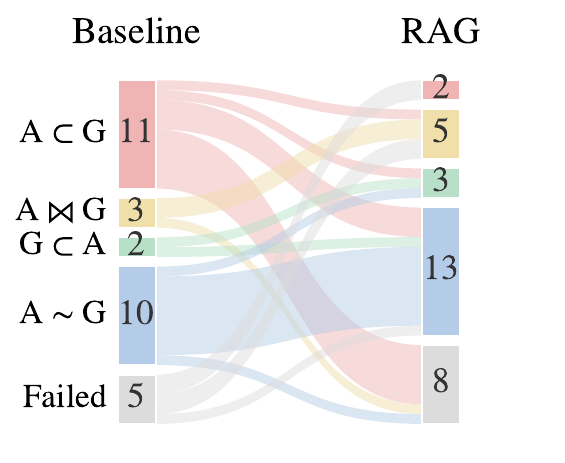}
    \caption{Kimi K2.5}
    \label{fig:rq3-kimi-rag}
  \end{subfigure}

  % 第二行：Baseline vs. History
  \begin{subfigure}[b]{0.24\textwidth}
    \includegraphics[width=\linewidth]{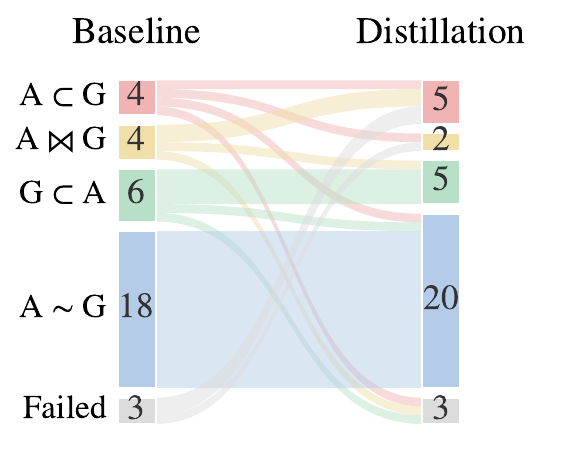}
    \caption{GPT-5.5}
    \label{fig:rq3-gpt-history}
  \end{subfigure}\hfill
  \begin{subfigure}[b]{0.24\textwidth}
    \includegraphics[width=\linewidth]{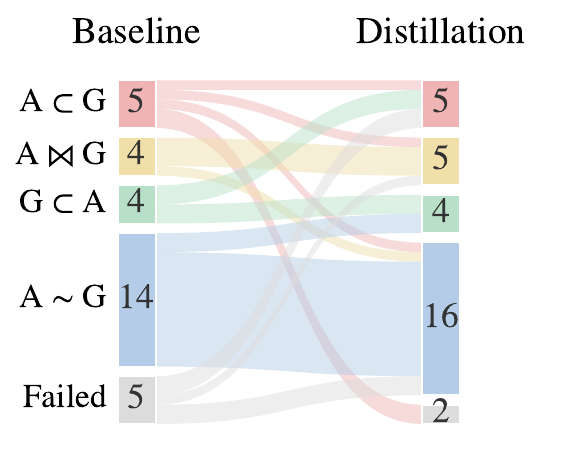}
    \caption{DeepSeek-V4-Pro}
    \label{fig:rq3-deepseek-history}
  \end{subfigure}\hfill
  \begin{subfigure}[b]{0.24\textwidth}
    \includegraphics[width=\linewidth]{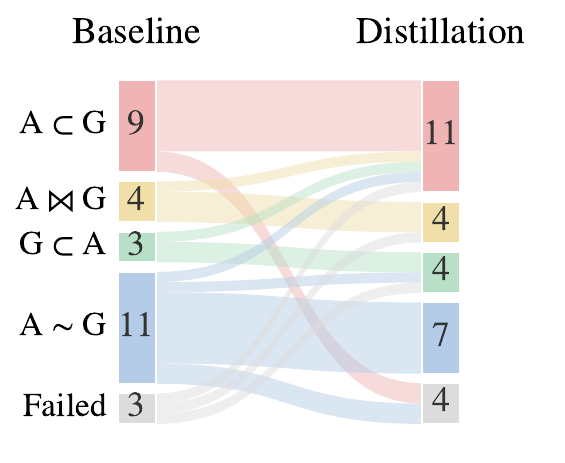}
    \caption{Qwen3.5-Plus}
    \label{fig:rq3-qwen-history}
  \end{subfigure}\hfill
  \begin{subfigure}[b]{0.24\textwidth}
    \includegraphics[width=\linewidth]{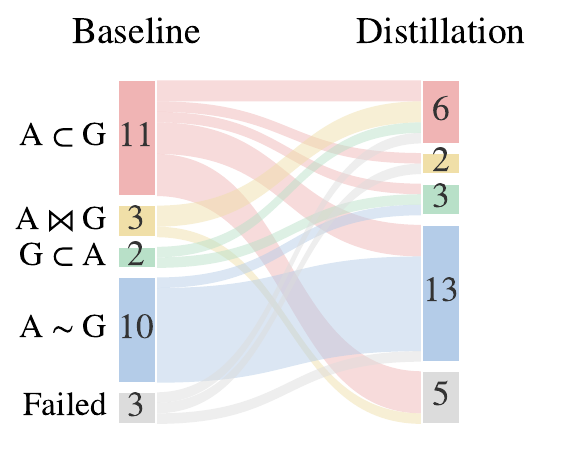}
    \caption{Kimi K2.5}
    \label{fig:rq3-kimi-history}
  \end{subfigure}

  \caption{Outcome transitions under baseline and different augmentation strategies. Issues whose generated patches fail to compile are classified as Failed. Issues that fail in both compared settings are omitted from the figure.}
  \label{fig:rq3}
  \vspace{-15pt}
\end{figure*}

\subsubsection{Distillation-Based Augmentation}
\label{sec:distillation-based}

% RAG-based augmentation introduces historical knowledge to the agent by retrieving a small set of issue-specific examples that are semantically related to the current missed-optimization case. In addition to this instance-level form of historical guidance, we also introduce a distillation-based augmentation method that presents historical knowledge in a more aggregated form. 
% Specifically, 
In addition to RAG,
we distill the generalization experience accumulated across historical optimization fixes into a compact knowledge base and provide it to the agent during patch generation.

The intuition is that historical developer knowledge can be represented at different levels of abstraction. Individual patches provide concrete examples of how a specific missed optimization was repaired, whereas recurring design choices across multiple patches reveal broader generalization practices. For example, developers may generalize an optimization from one predicate to a family of related predicates, from one operand order to both commuted forms, or from a single instruction shape to several semantically equivalent IR encodings. Distillation captures such cross-instance regularities and presents them as explicit background knowledge for the agent.

% \batu{it should be explained here that a one-time summary is a reasonable operation.}
To construct this knowledge, we aggregate historical generalization summaries by optimization pass and ask an LLM to synthesize pass-level knowledge documents. Fig.~\ref{fig:history} shows the workflow of distillation-based augmentation. Each document captures frequently missed elements, such as instruction forms, operand variants, predicates, type constraints, and analysis facts, as well as recurring developer generalization patterns, such as equivalent algebraic forms, multi-instruction canonicalizations, and interactions with value-tracking reasoning. During patch generation, we supply the pass-level distilled document as background guidance for agents.

% We evaluate distillation and RAG separately: RAG provides concrete issue-level precedents, while distillation provides summarized pass-level generalization principles. Both aim to help the agent infer the intended scope, variants, preconditions, and correctness constraints of the optimization.

% During patch generation, the pass-level knowledge document corresponding to the localized optimization pass is supplied to the agent as background guidance. 
% We evaluate distillation-based augmentation and RAG-based augmentation separately. RAG exposes historical knowledge through concrete, issue-level precedents, whereas distillation exposes it through summarized pass-level generalization principles. Both strategies aim to help the agent reason about the intended optimization scope, including the variants, preconditions, and correctness constraints considered in prior optimization work.

\subsection{Experimental Setup}

% We mainly describe the experimental settings for the augmentation module. We use DeepSeek-V4-Pro to generate the summary for each historical pull request. For Section~\ref{sec:rag-based}, we perform IR-pair matching on a single NVIDIA RTX A6000 GPU. Due to GPU memory constraints, we adopt Qwen3-Embedding-4B~\cite{zhang2025qwen3}, a well-known general-purpose embedding model, to compute embeddings for retrieval. For each test case under analysis, we retrieve the top three most similar summaries, i.e., \(k=3\). For Section~\ref{sec:distillation-based}, we use DeepSeek-V4-Pro to synthesize the distilled knowledge.

We focus on the experimental settings for the augmentation module. We use DeepSeek-V4-Pro to generate the summary for each historical pull request. DeepSeek-V4-Pro is also used for the one-time knowledge distillation in Section~\ref{sec:distillation-based}. This operation is practical because the input consists of compact historical summaries from past pull requests that are distinct from benchmark issues, where each summary is typically a few hundred tokens, and the full set of 869 summaries is well within the 1M-token context window of DeepSeek-V4-Pro. In addition, the distillation is performed offline and grouped by optimization pass, so each synthesis prompt contains only the summaries associated with the corresponding pass.

For RAG, we perform IR-pair matching on a single NVIDIA RTX A6000 GPU. Due to GPU memory constraints, we adopt Qwen3-Embedding-4B~\cite{zhang2025qwen3}, a well-known general-purpose embedding model, to compute embeddings for retrieval. For each test case under analysis, we retrieve the top three most similar summaries, i.e., \(k=3\). For distillation, the distilled pass-level knowledge is produced by DeepSeek-V4-Pro using the offline synthesis procedure described above.

\subsection{Empirical Findings}

Fig.~\ref{fig:rq3} shows how the scope relationship between agent-generated patches and golden patches changes after incorporating historical knowledge through RAG-based or distillation-based augmentation. We focus on generated patches that pass compilation and examine whether their optimization scopes exactly match or cover the developer-intended scope represented by the golden patch.

First, we observe consistent gains in exact scope alignment. For every base model, at least one historical-knowledge augmentation strategy increases the number of \(A \sim G\) cases over the non-augmented baseline. For example, GPT-5.5 improves from 18 exact matches to 22 with RAG and 20 with distillation, while Kimi K2.5 improves from 10 to 13 under both augmentation strategies. DeepSeek-V4-Pro also benefits from distillation, increasing from 14 to 16 exact matches, and Qwen3.5-Plus obtains a modest gain under RAG, increasing from 11 to 12. Moreover, the increase in \(A \sim G\) is not primarily explained by a reduction in \(G \subset A\). For instance, under RAG, GPT-5.5 increases \(A \sim G\) by four cases while \(G \subset A\) decreases by only three cases; Kimi K2.5 increases \(A \sim G\) by three cases while \(G \subset A\) also increases from 2 to 3. This suggests that historical knowledge helps agents correct scope decisions that would otherwise lead to misaligned patches, turning more originally non-aligned patches into exact scope matches.

Second, we examine the combined scope-covering outcome \(A \sim G \lor G \subset A\), which captures all compilable patches that cover the developer-intended scope. This combined category improves for several base models under at least one form of historical-knowledge augmentation. For GPT-5.5, scope coverage increases from 24 to 25 with RAG. DeepSeek-V4-Pro improves from 18 to 20 with distillation. Kimi K2.5 shows the largest gain, increasing from 12 to 16 under both RAG and distillation. These gains indicate that, beyond increasing exact matches, historical knowledge can also help agents generate patches that cover the developer-intended optimization scope.

Overall, historical-knowledge augmentation mainly improves agents' ability to infer the developer-intended generalization scope by increasing exact scope matches, while also improving broader scope coverage for several models. The gains are model-dependent, but the results consistently show that historical knowledge provides useful guidance for scope inference in missed-optimization repair.

\findingbox{Finding 3}{Historical knowledge improves agents' inference of developer-intended scope, increasing exact scope alignment and scope coverage for agents.}

\section{Real-World IR Optimization Evaluation}

% RQ3 indicates that historical knowledge improves agents' ability to decide and adjust optimization scopes, resulting in more exact matches with the golden patch and also more patches that cover the developer-intended scope. To validate whether these scope-level improvements translate into practical benefits, we further examine whether the resulting patches can trigger on IR patterns that occur in real-world downstream projects.

RQ3 indicates that historical knowledge improves agents’ ability to decide and adjust optimization scopes, resulting in more exact matches with the golden patch and also more patches that cover the developer-intended scope. We next examine whether these scope-level improvements also translate into greater real-world applicability on downstream IR.
We use optimization hits as a proxy for patch effectiveness because they capture whether a patch is exercised on real-world IR, which is a necessary precondition for producing any real-world optimization benefit. Although hits do not directly measure final performance, they provide a scalable and uniform signal across many local LLVM missed-optimization patches.

This study answers the following research question:
\begin{itemize}
    \item \textbf{RQ4:} Do the alignment improvements enabled by historical knowledge translate into changes in optimization hit behavior on real-world software?
\end{itemize}

\subsection{Experimental Setup}

We evaluate the real-world optimization effectiveness of patches by comparing the number of optimization hits they produce on downstream projects. For two patches addressing the same missed-optimization issue, we apply each patch to the same LLVM revision and build two corresponding LLVM compilers. We then use both compilers to process the same real-world project and record how many times the optimization introduced by the patch is triggered.

% We count an optimization hit whenever the patched optimization actually fires during compilation. Specifically, we instrument each patch so that a counter is incremented only after the optimization matches the target IR pattern, satisfies all required preconditions, and performs the corresponding rewrite. When a patch contains multiple modified code segments, we instrument the longest segment corresponding to the core optimization logic, which typically contains the final rewrite. Therefore, hit counts measure successful optimization applications rather than merely potential or syntactically similar IR patterns. Multiple firings within the same project are counted independently, and the project-level hit count is the sum of all firings across all IR files from that project.

We count an optimization hit only when the patched optimization actually fires. For patches with multiple modified segments, we instrument the longest segment containing the core rewrite logic because it typically corresponds to the unique semantic transformation implementing the optimization. Project-level hits sum all firings across the project's IR files, with repeated firings counted independently.

For each base model in the agent, we compare patches generated with historical knowledge against corresponding baseline patches. 
% We record success cases on each real-world project to evaluate whether historical knowledge leads to more realized optimization opportunities in practical IR workloads.
We use LLVM Opt Benchmark~\cite{opt-benchmark} as the source of real-world IR programs in this evaluation. This benchmark provides LLVM IR generated from a broad set of open-source projects. Since the IR is collected from real applications rather than synthetic test cases, it provides a suitable workload for measuring whether generated optimization patches can be exercised in practical compilation scenarios.
Due to the high cost of running all benchmark projects for every patch and model, we select eight representative projects from LLVM Opt Benchmark. The selected projects span C, C++, and Rust, and each project has more than 10K GitHub stars, indicating that they are widely used and actively recognized by open-source community. They are \texttt{git}, \texttt{linux}, \texttt{ffmpeg}, \texttt{opencv}, \texttt{llama.cpp}, \texttt{z3}, \texttt{uv}, and \texttt{tree-sitter}.

\subsection{Empirical Findings}

\begin{figure}[htbp]
    \centering
    \includegraphics[width=\linewidth]{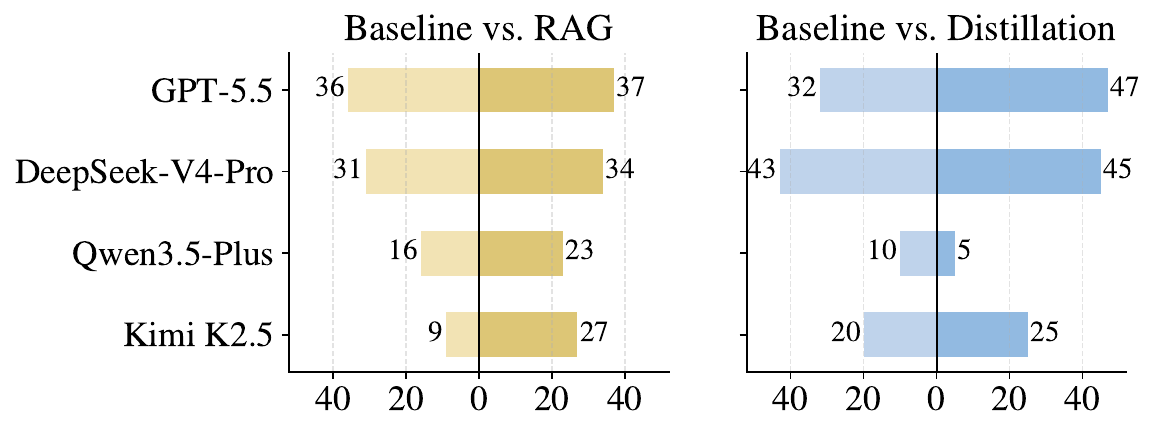}
    \caption{
Accumulated wins and losses of augmented patches against baseline patches over all evaluated issue-project pairs. 
% Each issue-project pair is classified by comparing the number of optimization hits produced by the augmented and baseline patches on the same project. 
}
    \label{fig:rq4_1}
\end{figure}

For each issue-project pair, we compare the historical-knowledge-augmented patch with the corresponding baseline patch. The augmented patch is counted as a win if it produces more optimization hits, a loss if it produces fewer hits, and a tie otherwise.
Fig.~\ref{fig:rq4_1} shows the accumulated wins and losses, omitting ties. In RAG setting, all four models achieve more wins than losses, suggesting that retrieved historical cases help patches trigger on more real-world IR patterns. Distillation also shows a positive trend, as three of the four models obtain more wins than losses, with GPT-5.5 showing the largest margin. Overall, historical knowledge improves the practical exposure of agent-generated optimizations on real-world IR, though the effect varies by model and augmentation strategy.

\vspace{-5pt}

\vspace{-5pt}

% We also report project-level cumulative patch-triggered hits in Fig.~\ref{fig:rq4_2}. For each benchmark project, we aggregate the total number of hits triggered by the baseline, RAG-augmented, and distillation-augmented patches across all four models and all evaluated issues. The distillation-augmented patches achieve higher cumulative coverage than the baseline on all evaluated projects, increasing the total number of hits from 20.2M to 24.3M. This improvement is observed across projects of different scales, including large codebases such as \texttt{opencv}, \texttt{uv}, and \texttt{linux}, as well as smaller projects such as \texttt{llama.cpp}. 

% The RAG-augmented patches also improve over the baseline in several projects, including \texttt{linux}, \texttt{uv}, and \texttt{tree-sitter}, resulting in 20.3M total hits overall. However, the gains are more project-dependent, with lower cumulative hits than baseline on some projects. Overall, these results show that historical augmentation can increase the number of optimization opportunities realized on real-world IR, and that the improvement depends on how effectively the generated patches generalize to  IR patterns present in each project.

\begin{figure}[t]
    \centering
    \includegraphics[width=\linewidth]{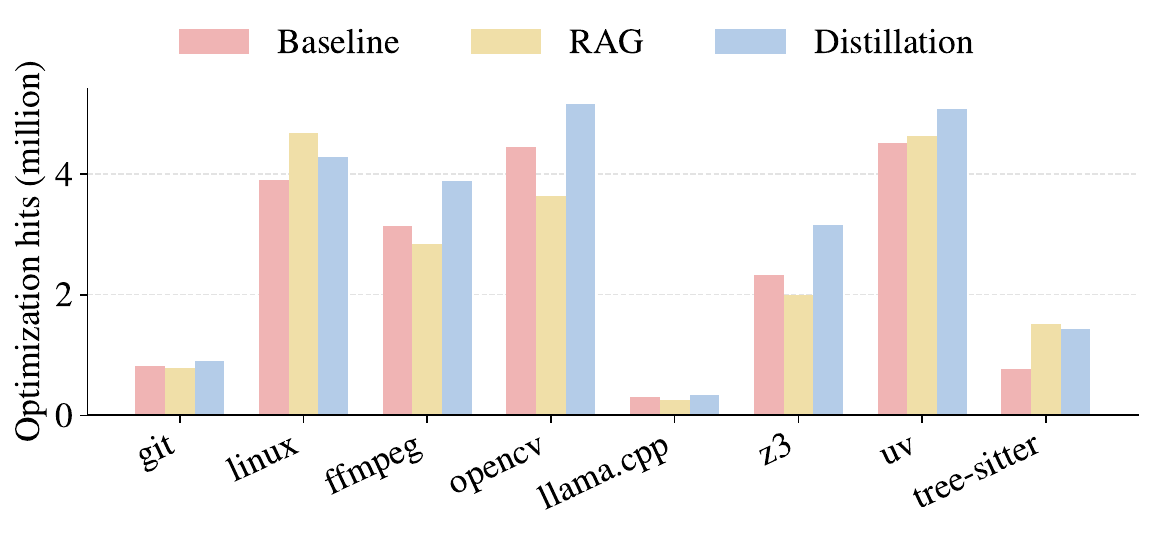}
    \caption{
Project-level cumulative optimization hits of baseline, RAG-augmented, 
and distillation-augmented patches across all evaluated issues and models.
}
    \label{fig:rq4_2}
    \vspace{-10pt}
\end{figure}

Fig.~\ref{fig:rq4_2} reports project-level cumulative hits aggregated across all models and issues. Distillation improves over the baseline on every evaluated project, increasing total hits from 20.2M to 24.3M across large codebases, such as \texttt{opencv}, \texttt{uv}, and \texttt{linux}, and smaller projects such as \texttt{llama.cpp}. RAG yields a smaller and more project-dependent gain, increasing total hits to 20.3M overall, with improvements on projects such as \texttt{linux}, \texttt{uv}, and \texttt{tree-sitter}. These results show that historical augmentation can increase realized optimization opportunities on real-world IR, but its effect depends on how the generated patches generalize to each project's IR patterns.

\findingbox{Finding 4}{Historical knowledge increases optimization-hit coverage on real-world IR, providing evidence that the generated patches are exercised more frequently in practice.}

\section{Case Study}
\label{sec:case-study}

\begin{figure*}[t]
    \centering

    \begin{minipage}[t]{0.32\textwidth}
        \centering
\begin{lstlisting}[style=irstyle, basicstyle=\scriptsize\ttfamily]
define i1 @src(i8 %a) {
  %cmp1 = icmp ult i8 %a, 5
  %masked = and i8 %a, -2
  %cmp2 = icmp eq i8 %masked, 2
  %and = and i1 %cmp1, %cmp2
  ret i1 %and
}

define i1 @tgt(i8 %a) {
  %masked = and i8 %a, -2
  %cmp2 = icmp eq i8 %masked, 2
  ret i1 %cmp2
}
\end{lstlisting}
        \caption{The issue reports a masked equality check implies a range check.}
        \label{fig:case-158326-ir}
    \end{minipage}
    \hfill
    \begin{minipage}[t]{0.32\textwidth}
        \centering
\begin{lstlisting}[style=irstyle, basicstyle=\scriptsize\ttfamily]
define i1 @src(i8 %a) {
  %cmp1 = icmp ult i8 %a, 3
  %masked = and i8 %a, -2
  %cmp2 = icmp eq i8 %masked, 2
  %and = and i1 %cmp1, %cmp2
  ret i1 %and
}

define i1 @tgt(i8 %a) 
{
  %and = icmp eq i8 %a, 2  
  ret i1 %and  
}

\end{lstlisting}
        \caption{A golden test missed by the direct-repair patch.}
        \label{fig:case-158326-direct-fail}
    \end{minipage}
    \hfill
    \begin{minipage}[t]{0.32\textwidth}
        \centering
\begin{lstlisting}[style=irstyle, basicstyle=\scriptsize\ttfamily]
define i1 @src(i8 %a) {
  %cmp1 = icmp ult i8 %a, 5
  %masked = and i8 %a, -3
  %cmp2 = icmp eq i8 %masked, 1
  %and = and i1 %cmp1, %cmp2
  ret i1 %and
}

define i1 @tgt(i8 %a) {
  %masked = and i8 %a, -3
  %cmp2 = icmp eq i8 %masked, 1
  ret i1 %cmp2
}
\end{lstlisting}
        \caption{An LLM-generated test that golden patch does not handle.}
        \label{fig:case-158326-generalization-fuzz}
    \end{minipage}

\end{figure*}
  % \vspace{-15pt}

\begin{figure*}[t]
\centering
\begin{minipage}[t]{0.31\textwidth}
\vspace{0pt}
\begin{casebox}[1. Retrieved context]{ragblue}
\begin{lstlisting}[style=casecode,language=LLVMIR]
define i1 @src(i32 %x) {
  %and = and i32 %x, -8
  %cmp = icmp ult i32 %and, 1
  ret i1 %cmp
}
define i1 @tgt(i32 %x) {
  %cmp = icmp ult i32 %x, 8
  ret i1 %cmp
}
\end{lstlisting}
\footnotesize
The key idea is to reason about the value range
represented by the masked expression.
\end{casebox}
\end{minipage}
\hfill
\begin{minipage}[t]{0.31\textwidth}
\vspace{0pt}
\begin{casebox}[2. Agent reasoning]{reasonorange}
\begin{lstlisting}[style=casecode]
icmp ult x, 5        -> x in [0, 5)
icmp eq (x & -2), 2 -> x in [2, 4)
\end{lstlisting}

{\footnotesize\raggedright
Since [2, 4) lies inside [0, 5),\\
the range check is redundant.\\
Extend \texttt{foldAndOrOfICmps-}\\
\texttt{UsingRanges} to:\\
\hspace*{0.8em}look through \texttt{(x \& -Pow2)},\\
\hspace*{0.8em}derive the masked range,\\
\hspace*{0.8em}merge it with the icmp range.\\
\par}
\end{casebox}
\end{minipage}
\hfill
\begin{minipage}[t]{0.31\textwidth}
\vspace{0pt}
\begin{casebox}[3. Patch abstraction]{patchgreen}
\begin{lstlisting}[style=casecode]
if (icmp is equality over
    (X & -Pow2)) {
  maskedRange =
    getRangeForMaskedICmp(...);
} else {
  icmpRange =
    makeExactICmpRegion(...);
}
range=merge(maskedRange,icmpRange);
foldByRange(range);
\end{lstlisting}
\footnotesize
Combine ranges by
intersection or union.

\end{casebox}
\end{minipage}
\caption{How RAG guides the agent from retrieved masked-comparison knowledge to range-based patch generation. 
The retrieved example suggests a range interpretation for masked comparisons; the agent applies this abstraction to the issue-level pattern and integrates it into the existing range-combination logic.}
\label{fig:case_rag_flow}
\vspace{-15pt}
\end{figure*}

We use LLVM issue~\#158326 as an example. As shown in Fig.~\ref{fig:case-158326-ir}, the issue reports a range check combined with a masked equality check: \texttt{(a \& -2) == 2} restricts \texttt{a} to \texttt{\{2,3\}}, making \texttt{a < 5} redundant. More generally, the intended pattern is:
% \((X \ \mathit{pred}\ N) \ \mathit{op}\ ((X \& -2^k) = C)\), where \(\mathit{op} \in \{\land,\lor\}\). 
\vspace{-10pt}
\begin{equation}
\label{eq:pred-masked-pattern}
(X \ \mathit{pred}\ N) \ \mathit{op}\ ((X \mathbin{\&} -2^k) = C),
\quad
\mathit{op} \in \{\land,\lor\}.
\end{equation}
\vspace{-15pt}

A direct patch can pass the issue-level test, but the golden patch targets a broader range-reasoning abstraction by converting both ordinary comparisons and masked equalities into \texttt{ConstantRange} objects and combining them through intersection or union in \texttt{foldAndOrOfICmpsUsingRanges}.

With DeepSeek-V4-Pro, both direct-repair patch and generic generalization instruction-guided patch pass initial test but fail 3 of 5 golden tests.
Fig.~\ref{fig:case-158326-direct-fail} shows one missed golden case: \texttt{(a \& -2) == 2} gives \(\texttt{a} \in \{2,3\}\), while \texttt{a < 3} gives \(\texttt{a} \in \{0,1,2\}\); their intersection is \(\{2\}\), so the conjunction should simplify to \texttt{a == 2}. The direct patch does not perform this range intersection. Moreover, generic generalization instruction-guided patch expands syntactically to related masks, as illustrated by LLM-generated test in Fig.~\ref{fig:case-158326-generalization-fuzz}, but still fails to learn the intended semantic range abstraction.

The retrieved examples provide evidence for why the RAG run behaves differently. As shown in Fig.~\ref{fig:case_rag_flow}, the retrieved context does not reveal the target patch directly. Instead, it exposes a related InstCombine pattern where comparisons over values masked by a negated power-of-two mask can be understood as range constraints over the original value. This context is relevant to the issue because the failing program also combines a range check with a masked comparison.

% \begin{figure}[t]
%     \centering
% \begin{lstlisting}[style=irstyle, xleftmargin=1em]
% Retrieved IR:
% define i1 @src(i32 %x) {
%   %and = and i32 %x, -8
%   %cmp = icmp ult i32 %and, 1
%   ret i1 %cmp
% }
% define i1 @tgt(i32 %add) {
%   %cmp = icmp ult i32 %add, 8
%   ret i1 %cmp
% }

% Retrieved summary:
%   The developer generalized the pattern
%     (and X, highmask) pred C
%   where highmask is a negated power-of-two mask.
%   The key idea is to reason about the value range
%   represented by the masked expression.
% \end{lstlisting}
%     \caption{Excerpt of RAG-retrieved context provided to the agent.}
%     \label{fig:case-158326-rag-context}
% \end{figure}

% The agent's subsequent reasoning follows this retrieved clue by lifting the issue-specific implication into a range-level formulation. As shown in Fig.~\ref{fig:case_rag_flow}, the agent identifies that the masked equality is not merely a syntactic pattern, but a constraint over the original variable. For example, \texttt{(x \& -2) == 2} corresponds to the range \([2,4)\), which is contained in the range required by \texttt{x < 5}. This leads the agent to \texttt{foldAndOrOfICmpsUsingRanges}, the existing InstCombine implementation point for combining comparisons via \texttt{ConstantRange}.

Following retrieved clue, the agent lifts issue-specific implication into a range-level formulation. As shown in Fig.~\ref{fig:case_rag_flow}, it treats masked equality as a constraint over the original variable rather than as a purely syntactic pattern. For example, \texttt{(x \& -2) == 2} denotes the range \([2,4)\), which is contained in the range required by \texttt{x < 5}. This reasoning points the agent to \texttt{foldAndOrOfICmpsUsingRanges}, existing InstCombine logic for combining comparisons through {ConstantRange}.
The RAG patch extends range-folding logic by converting masked equality over \texttt{x \& -Pow2} into a {ConstantRange} over \texttt{x}, allowing existing range-combination logic to simplify the operation, as shown in Fig.~\ref{fig:case_rag_flow}.

This case illustrates our central finding. The issue contains a masked-equality pattern implying a range check, while direct fixes and generic generalization miss broader predicate combinations. RAG retrieves a related masked-comparison example, prompting range-based reasoning. The resulting patch passes all golden tests, yields no accepted counterexamples, and is therefore considered as behaviorally indistinguishable.

\section{Related Work}

% \subsection{Coding Agent for Issue Fixing}
% Recent research~\cite{yang2024sweagent,li2025patchpilot} and engineering practice~\cite{anthropic_claude_code,openai_codex}
% have increasingly focused on developing coding agents that can generate patches
% to fix issues across a wide range of software repositories serving diverse purposes
% and application domains. This growing interest is not limited to such
% general-purpose repository settings, but has also expanded into specialized domains
% such as compilers. Zheng et al.~\cite{zheng2026agentic} propose llvm-autofix,
% a compiler-specific agentic harness for automating LLVM bug repair. It combines
% LLVM-tailored tools, a reproducible benchmark, and a lightweight repair agent
% to better support LLM-based compiler debugging. Compared with llvm-autofix, which
% primarily focuses on fixing miscompilations in LLVM, our work targets missed
% optimization issues in LLVM and specifically studies generalization, a unique
% challenge encountered when fixing this type of issue.

\textbf{Agents for Issue Fixing.}
Recent work has explored coding agents for patch generation across diverse repositories~\cite{yang2024sweagent,li2025patchpilot,anthropic_claude_code,openai_codex}, with increasing attention to compiler repair. Zheng et al.~\cite{zheng2026agentic} propose llvm-autofix, an LLVM-focused agentic harness for bug repair. In contrast, our work targets missed optimizations and their distinctive generalization challenges.

% Recent work~\cite{yang2024sweagent,li2025patchpilot} and practice~\cite{anthropic_claude_code,openai_codex} have explored coding agents for patch generation across diverse repositories, with growing interest in specialized domains such as compilers. Zheng et al.~\cite{zheng2026agentic} propose llvm-autofix, a compiler-specific agentic harness for LLVM bug repair with tailored tools, a reproducible benchmark, and a lightweight repair agent. Unlike llvm-autofix, which mainly targets LLVM miscompilations, our work focuses on missed optimization issues and studies their distinctive generalization challenge.

\textbf{Generalizing Optimization.}
Generalization has long been studied in compiler. Prior work on peephole-optimization generalization~\cite{buchwald2015optgen,tate2010generating} explored symbolic enumeration and proof-centric methods, while Hydra~\cite{mukherjee2024hydra} and LPG~\cite{liao2026leveraging} improve scalability through generate-and-verify workflows and LLM-assisted generalization from real-world optimizations. In contrast, our work unifies generalization and patch generation by starting from an issue report, inferring the intended generalization, and directly producing the corresponding patch.

\textbf{Developer-Aligned Patch.}
Prior work distinguishes plausible patches that pass tests from correct patches that reflect developer intent, showing that test-passing patches can still overfit or be incorrect~\cite{smith2015cure,qi2015analysis,xiong2018identifying}. This has motivated using historical fixes as references for intended behavior~\cite{kim2013automatic,long2016automatic,le2016history,bader2019getafix}. We extend this idea to compiler missed optimizations, where correctness means matching the developer-intended optimization scope rather than merely passing tests.

\section{Discussion}

% 还有就是数据泄露问题，说patch和正式修复的patch长得很不一样
% 没有验证fuzzer的有效性，也就是fuzz不出来不一定代表没问题
% 为什么只在一个agent上做实验
% 重复性问题？

\textbf{Potential pretraining-data leakage.}
Since LLVM patches are public, base models may have seen some benchmark fixes during training. However, generated patches are usually not textually close to the golden patches and often differ in implementation strategy or code location, suggesting that successes are unlikely to be simple memorization. 

\textbf{Stochasticity and evaluation design.}
We use an open and controllable agent harness rather than productized coding agents with hidden prompts, tools, policies, or updates, ensuring a fixed interaction protocol across all experiments. While agentic repair exhibits stochasticity due to LLM sampling and tool-use ordering, we evaluate each issue–configuration pair under a fixed execution budget and record a single representative outcome to enable consistent issue-level comparison and transition-based analysis. This design focuses on measuring systematic shifts induced by different augmentation strategies rather than run-to-run variance. The observed trends are consistent across four base models and two augmentation strategies and are further corroborated by real-world results.

\section{Conclusion}

% We studied agent-based patching for LLVM missed-optimization issues, focusing on whether generated patches match developers' intended generalization scope. Our results show that agents often fix reported examples but frequently under-generalize or over-generalize relative to golden patches, while historical knowledge from prior LLVM optimization pull requests, provided through retrieval or distillation, improves scope alignment and can increase optimization hits on real-world IR. Overall, effective automated compiler patching requires not only strong code generation models, but also domain-specific historical knowledge that helps transform concrete missed-optimization reports into semantically well-scoped patches.

We studied agent-based patching for LLVM missed-optimization issues, focusing on whether generated patches match developers’ intended generalization scope. Results show agents often fix reported examples but under- or over-generalize relative to golden patches. Historical knowledge from prior LLVM optimization PRs, via retrieval or distillation, improves scope alignment and can increase real-world IR optimization hits. Overall, effective automated compiler patching requires domain-specific historical knowledge to turn concrete reports into semantically well-scoped patches.

\section*{Data Availability}
Our research artifacts are publicly available at:
\url{https://github.com/cuhk-s3/understanding-generalization}.
\bibliographystyle{IEEEtran}
\bibliography{refs}

\end{document}